\documentclass[fleqn,usenatbib]{mnras}

\usepackage{newtxtext,newtxmath}

\usepackage[T1]{fontenc}

\DeclareRobustCommand{\VAN}[3]{#2}
\let\VANthebibliography\thebibliography
\def\thebibliography{\DeclareRobustCommand{\VAN}[3]{##3}\VANthebibliography}

\usepackage{subcaption}
\usepackage{graphicx}	
\usepackage{amsmath}	
\usepackage{xcolor}
\usepackage{multirow}

\author[Singha et al.]{\parbox{\textwidth}{
Jaikhomba Singha$^{1,2}$,\thanks{E-mail:jaikhomba.singha@uct.ac.za}
Bhal Chandra Joshi$^{3,2}$, 
M. A. Krishnakumar$^{4, 5, 3}$, 
Fazal Kareem$^{6,7}$, 
Adarsh Bathula$^{8}$, 
Churchil Dwivedi$^{9}$, 
Shebin Jose Jacob$^{10}$, 
Shantanu Desai$^{11}$,
Pratik Tarafdar$^{12}$, 
P. Arumugam$^{2}$, 
Swetha Arumugam$^{13}$, 
Manjari Bagchi$^{12,14}$,
Neelam Dhanda Batra$^{15}$, 
Subhajit Dandapat$^{16}$, 
Debabrata Deb$^{12}$, 
Jyotijwal Debnath$^{12,14}$,
A Gopakumar$^{16}$, 
Yashwant Gupta$^{3}$, 
Shinnosuke Hisano$^{17}$, 
Ryo Kato$^{18, 19}$,
Tomonosuke Kikunaga$^{20,21}$,
Piyush Marmat$^{2}$,
K. Nobleson$^{17}$, 
Avinash K. Paladi$^{22}$,  
Arul Pandian B.$^{23}$, 
Thiagaraj Prabu$^{23}$,
Prerna Rana$^{16}$,
Aman Srivastava$^{11}$,
Mayuresh Surnis$^{24}$,
Abhimanyu Susobhanan$^{25}$,
Keitaro Takahashi$^{19,17}$\newline
\emph{\normalsize Affiliations are listed at the end of the paper.}
}}

\title[Improving DM estimates using low-frequency scattering-broadening estimates]{Improving DM estimates using low-frequency scattering-broadening estimates}
\date{Accepted XXX. Received YYY; in original form ZZZ}

\pubyear{2021}

\begin{document}
\label{firstpage}
\pagerange{\pageref{firstpage}--\pageref{lastpage}}
\maketitle


\begin{abstract}
A pulsar's pulse profile gets broadened at low frequencies due to dispersion along the line of sight or due to multi-path propagation. The dynamic nature of the interstellar medium  makes both of these effects time-dependent and introduces slowly varying time delays in 
the measured times-of-arrival similar to those introduced by passing gravitational waves. 
In this article, we present an improved method to correct for such delays by obtaining unbiased dispersion measure (DM) measurements by  
using low-frequency estimates of the scattering parameters.
We evaluate this method by comparing the obtained DM estimates with those, where scatter-broadening is 
ignored using simulated data. A bias is seen in the estimated DMs for simulated data with pulse-broadening 
with a larger variability for a data set with a variable frequency scaling index, $\alpha$, as compared 
to that assuming a Kolmogorov turbulence. Application of the proposed method removes this bias robustly 
for data with band averaged signal-to-noise ratio larger than 100.  We report the measurements of the scatter-broadening 
time and  $\alpha$ from analysis of 
PSR J1643$-$1224, observed with upgraded Giant Metrewave Radio Telescope as part of the 
Indian Pulsar Timing Array experiment. These scattering parameters were found to vary with epoch and $\alpha$ was different from that 
expected for Kolmogorov turbulence. Finally, we present the DM time-series after application of this technique to PSR J1643$-$1224.
\end{abstract}




\section{Introduction}  \label{sec:intro}
The precision in the time of arrival (ToA) of a pulsar's radio pulse is determined in part by how bright and sharp the received pulse is. Both of these quantities, namely the signal-to-noise ratio (S/N) and the pulse width, are affected by the propagation of the pulsed signal through the ionised interstellar medium (IISM). The IISM can impose a frequency-dependent delay on the pulses, which, when added together without proper correction, will make the pulse appear smeared. This dispersion is mainly caused by the integrated column density of electrons along the line of sight and is quantified by the Dispersion Measure (DM). In addition, electron density inhomogeneities in the IISM encountered along the line of sight lead to multi-path propagation of radio waves, which also broadens the pulse \citep{Rickett_1977}. This pulse broadening can be mathematically described as a convolution of the intrinsic pulse profile with a pulse broadening function, such as $\exp(-\phi/\tau_{sc})$, where $\phi$ is the pulse phase and $\tau_{sc}$ is the scatter-broadening time scale in the case of a thin scattering screen \citep{Williamson1972}. Different  methods have been proposed in literature in order to obtain the scatter-broadening time scales. Several fitting techniques have been used \citep{Lohmer2001, Lohmer2004, Krishnakumar_2015, Krishnakumar_2017, Geyer2017, Krishnakumar_2019} to estimate pulse broadening parameters for a sample of pulsars. Multiple works \citep{Bhat2004, Kirsten2019, Young2024} have used techniques based on the \texttt{CLEAN} \citep{Hogbom1974, bcc03} algorithm. The scatter-broadening time scales can also be estimated using scintillation bandwidth \citep{Cordes1985}. A complementary method uses cyclic spectroscopy (CS) \citep{Demorest2011} to determine the impulse response function of the ISM and thereby estimating pulse broadening times \citep{Walker2013}.

Both the phenomena, scattering and dispersion, are time-variable due to the dynamic nature of IISM. This variation induces a slowly varying chromatic time delay in the ToA measurements. The timescale of this stochastic delay is similar to that of the gravitational wave (GW) signature arising from an isotropic stochastic gravitational wave background (SGWB) formed by the random superposition of GWs emitted by an ensemble of super-massive black hole binaries \citep{Burke2019}. Hence, the wrong characterisation of this chromatic delay, or the individual pulsar chromatic noise, can lead to the false detection of SGWB \citep{Zic_PPTA_2022}. 

The measurement and characterization of this IISM noise is therefore crucial for experiments, which use a collection of pulsars to observe the GW signal from SGWB~\citep{Srivastava23}. These experiments are called pulsar timing arrays (PTAs).  There are four PTAs, which pool their data as part of the International 
Pulsar Timing Array  consortium \citep[IPTA : ][]{Hobbs2010,verbiest_et_al2016} : the European Pulsar Timing Array \citep[EPTA : ][]{Desvignes_2016, kc2013}, the Indo-Japanese pulsar timing array 
\citep[InPTA : ][]{JoshiAB+18,joshi_et_al2022,inptadr1}, 
the North American Nanohertz Observatory for Gravitational Waves \citep[NANOGrav : ][]{McLaughlin2013} 
and the Parkes Pulsar Timing Array \citep[PPTA : ][]{mhb+2013}. Recently, the MeerKAT Pulsar Timing Array
\citep[MPTA : ][]{MPTA_2023, bja+2020} and the Chinese Pulsar
Timing Array \citep[CPTA: ][]{Lee2016} have also started pulsar timing experiments. 

The estimates of DM in these PTA experiments are usually obtained by quasi-simultaneous/simultaneous or even observations separated by few days, at two or three different observing frequencies~\citep{Arzoumanian_2018, inptadr1}. The alignment of the fiducial point of the pulse at different observing frequencies is critical in such measurements. The scatter-broadening can introduce a systematic phase shift in the pulse's fiducial point. In the measurement procedure, this needs to be accounted for to avoid a systematic bias in the measured DMs. 
Furthermore, slow variations in  $\tau_{sc}$ over long periods of time can introduce corresponding variations in the measured DM values. Lastly, timing events, such as the ones reported in PSR J1713+0747 \citep{LamEG+2018, Goncharov+2020, j1713-1}, produce a discontinuity in the Gaussian process DM models, if accompanied by changes in $\tau_{sc}$. These epoch dependent systematic biases in the  DM estimates induce time varying delays in the ToAs, 
which act as a chromatic noise to SGWB signal. 
This noise, introduced by scatter-broadening variations,  needs to be  accounted for a reliable characterisation of the SGWB signal in PTA experiments. 
The correction of scatter-broadening in order to obtain robust estimates of DMs, and removal of this noise is the primary motivation of this study.

A few attempts have been reported in literature to mitigate the effects of scatter broadening from pulsar timing data. Particularly, in \cite{Mckee2018} the scattering information has been utilised to modify the template and obtain an additional delay in ToAs arising from scatter broadening. These delays were estimated by the difference 
between the centroid  of the scattered profile and that of the unscattered template. These delays  were further used to obtain the required corrections to DMs. No reconstruction of a profile without scatter-broadening was attempted in this method. Other approaches have also been proposed to increase the ToA precision in the  presence of scattering \citep{Levin2016, Lentati2017}, but these methods did not focus particularly on DM estimations. While the reconstruction of profiles 
without scattering from scattered profiles have been attempted using  techniques based on the  \texttt{CLEAN} \citep{Hogbom1974, bcc03} algorithm \citep{Bhat2004}, such reconstructed profiles were not used to estimate DM. Thus, our 
approach is therefore different from these methods in the literature, as we obtain unbiased DMs using reconstructed profiles in this paper.

The characterisation of scatter-broadening noise can be achieved with wide-band observations of millisecond pulsars (MSPs). Recently, wide-band receivers have been employed by the uGMRT \citep[ 300-500 MHz : ][]{ugmrt,inptadr1}, Parkes radio telescope \citep[800 $-$ 5000 MHz]{hobbs_2020_UWBP}, and  CHIME \citep[400 $-$ 800 MHz]{Amiri_2021} for higher precision DM measurements. The scatter-broadening noise can be well characterised with such wide-band receivers.  
However, the dispersive delay due to the IISM varies as $f^{-2}$, whereas the pulse scatter-broadening evolves as $f^{-4.4}$ if a Kolmogorov turbulence is assumed in the IISM, where $f$ is the observational frequency \citep{Rickett_1977}. This makes these propagation effects dominant at frequencies below 800 MHz {\citep{Lam_2016}, necessitating low-frequency measurements. While the DMs for nearby pulsars can be measured accurately with telescopes operating at very low frequencies ~\citep{Donner2020LoFAR, Bondonneau2021NenuFAR}, robust and precision DM measurements for moderately high DM pulsars, which are heavily scattered at very low frequencies, are only possible if the scatter-broadening variations estimated from such observations can be removed from the data. In this paper, we present an improved technique to achieve this and evaluate the efficacy of this technique using simulated data as well as data on a pulsar with significant pulse broadening. The various techniques available in literature to mitigate scatter-broadening and/or estimate DMs, elucidated in previous paragraphs, were implemented with narrow frequency bands, whereas our method utilises wideband low frequency data collected using the uGMRT.

The paper is arranged as follows. An improved technique to remove the 
effect of pulse scatter-broadening is described in Section \ref{sec:technique}. 
The technique was tested first with simulated data with a known injection 
of DM and scatter-broadening variations, and the results are 
presented in Section \ref{sec:simulations_test}. Results obtained by applying the technique on the InPTA data for PSR J1643$-$1224 
are discussed in Section~\ref{sec:result_1643} followed by our conclusions in Section~\ref{sec:conclusion}.

\begin{figure*}
 \centering
\includegraphics[scale=0.6,trim={0.1cm 0.1cm 0.1cm 0.1cm},clip]{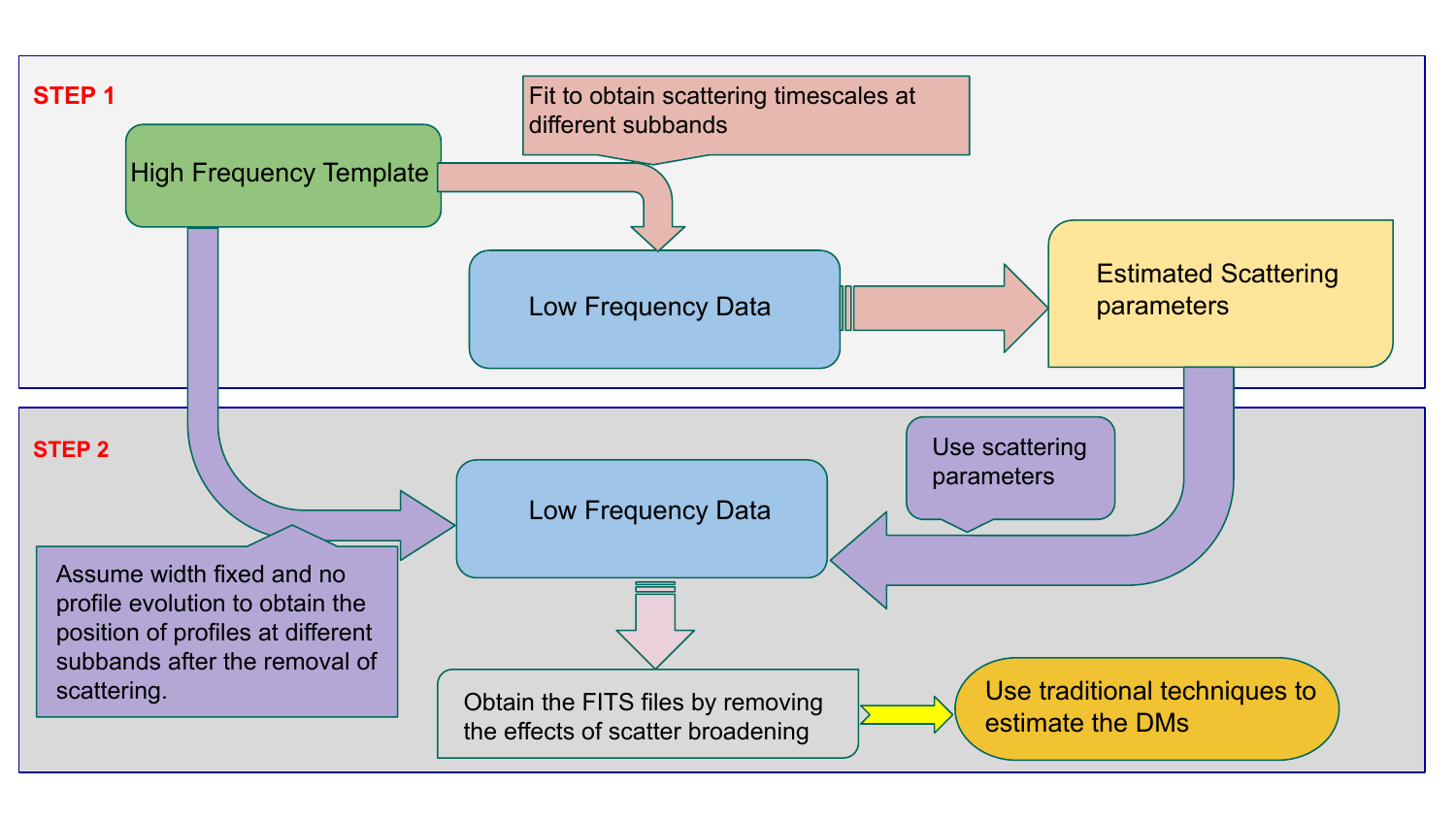}
\caption{Schematic diagram of the technique for obtaining pulse profile by removing scatter-broadening using low-frequency observations.} 
\label{dmscatblock}
\end{figure*}

\section{Description of the technique}
\label{sec:technique}

In this section, we describe in detail the technique used in this paper. There are two major steps involved in this technique. In the first step, we estimate the scattering parameters of the pulsar using low frequency data (between 300$-$500 MHz) by employing the method described in \cite{Krishnakumar_2015, Krishnakumar_2017, Krishnakumar_2019}. In the next step, we make use of these measurements of pulse broadening to recover the pulse shape, free from scattering. The aim of the technique is to recover these frequency resolved pulse profiles and then 
use them to obtain estimates of DM. The procedure used in the technique 
is shown schematically in Figure~\ref{dmscatblock}. 

\subsection{Estimation of scattering parameters}
The pulse broadening measurements were obtained as follows. We use the frequency-resolved integrated pulse profiles with a chosen number of sub-bands between 300$-$500 MHz. The number of 
sub-bands were selected to obtain a pulse profile with S/N of at least 50  in each sub-band. 
Then, a template profile is generated from a high S/N pulse profile by collapsing the data at 1260$-$1460 MHz, where the pulse broadening is negligible. Next, this template is convolved with a pulse broadening function, $\exp({-\phi/\tau_{sc}})$. The convolved template  is 
given by: 
\begin{equation}
\mathcal{F}(\phi) = a \times s(\phi - b) * \exp(-\phi/\tau_{sc}) \, ,
\end{equation}
where $s(\phi)$ is a high frequency template with amplitude $a$, $\phi$ is the pulse phase with peak at phase $b$, $\tau_{sc}$ is the pulse broadening time scale and $*$ denotes convolution. $\mathcal{F}(\phi)$ is then fitted to the observed pulse profile at each sub-band, keeping $\tau_{sc}$ as a fitted parameter, by minimizing the sum-of-squares of residuals obtained by subtracting the observed profile from $\mathcal{F}(\phi)$. This fit is carried out for each sub-band between 300$-$500 MHz data obtained using the InPTA observations and provides measurements of $\tau_{sc}$ as a function of observing frequency. The estimated $\tau_{sc}$ is then fitted to a power law model of the following form: 
\begin{equation}
\tau_{sc}(f) = \tau_0  f^{\alpha} 
\end{equation}
Here, $\tau_0$ is the pulse broadening at a reference frequency (e.g., 300 MHz) and $\alpha$ is the frequency scaling index of the scattering medium. This fit provides a measurement of $\alpha$ for each epoch. 

\subsection{Reconstruction of low frequency profile without scattering}
These $\alpha$ measurements can be now used to reconstruct the pulse 
profiles without scatter-broadening and thereby obtain the pulse phase 
(pulse position) corrected for scattering. A fit to these pulse phases 
across the band therefore provide more reliable measurements of DM. This is the improved technique presented in this paper, where
we use the same high-frequency template convolved with the scattering function, $\exp({-\phi/\tau_{sc}})$ but this time with the values of $\tau_{sc}$ estimated from the previous step to obtain a convolved profile, $T$. The sum-of-squared difference between the convolved 
profile and the observed scatter-broadened profile is given by: 
\begin{equation}
\mathcal{R}^2 = \sum (P_i - T_i)^2 \, ,
\end{equation}
where $P_i$ and $T_i$ are the $i$-th bin amplitudes of the observed scatter-broadened profile and convolved profile respectively.  $R^2$ is minimized (least square minimization), keeping 
$\tau_{sc}$ fixed to the parameter estimated in the previous step and allowing the amplitude ($a$) and peak position ($b$) of the convolved pulse profile to vary. The residuals after the fitting are given by: 
\begin{equation}
R_i = P_i - T_i .
\end{equation}
For a good fit, $R_i$ is normally distributed and represents the noise in the profile.  

Thus, the template profile, $s_i$, scaled by the amplitude at the fitted position provides a good representation of the pulse profile without scatter-broadening.  
This method is applied to all the sub-bands.
and the obtained profiles 
are written back to a  new PSRFITS file after adding the residuals, $R_i$ (noise), for each of the sub-bands. These profiles can now be used for estimating the DMs with conventional methods. While cyclic spectroscopy \citep{Demorest2011}, can be used to reconstruct the pulse profiles \citep{Walker2013}, our method of reconstruction of low frequency profiles free from scattering is unique. While in ~\citet{Mckee2018}, the scattering measurements were used to modify the \texttt{template} for a particular window (of 15 days) and estimate delay corrections in the time of arrival of the pulses from the difference in the centroids of the template and scattered profile, and consequently the corrections required to the DM values, our technique differs  from this in the sense that we are using the scattering measurements from individual epoch to reconstruct scatter free low frequency profiles 
of the relevant epoch and estimating its DM from the reconstructed profiles.

It is important to note that the main 
assumption in these steps is that the profile of the pulsar does not evolve significantly with frequency. This may not hold true for most pulsars. However, a few of the MSPs monitored by PTAs do not show profile evolution with frequency (eg. J1643$-$1224, J1909$-$3744, J1744$-$1134, etc.)~\citep{Dai2015}.

\section{Tests on Simulated Data}
\label{sec:simulations_test}   
\subsection{Simulations}
We simulated frequency-resolved PSRFITS \citep{Hotan+2004} files using the parameter file of PSR J1643$-$1224  
obtained from InPTA DR1 \citep{inptadr1}.  
The primary objectives of our simulations were:
\begin{enumerate}
    \item To gain an understanding of the impact of scatter-broadening on the DM estimation. Here, we explored two scenarios: one involved  a scattering process characterised by the Kolmogorov turbulence spectrum ($\alpha = -4.4$), and the other involved 
    a scattering process with varying  $\alpha$.
    
    \item To validate and assess the efficacy of the \texttt{DMscat} software.
\end{enumerate}

First, a single component pulse profile was simulated by generating a Gaussian placed at 
the middle of the pulse phase with a chosen width. For a given S/N across the band, 
the root mean square (RMS) of the required normally distributed noise was obtained 
by dividing the area under the pulse by the required S/N adjusted by the 
number of sub-bands. Noise with this RMS was then generated from a random 
number generator. This noise was added to each sub-band profile after 
convolving the pulse with the scatter-broadening function as described below. 
Data were simulated  with S/N varying between 10 to 2000 (10, 20, 30, 50, 100, 400 and 
2000).

We assumed a thin-screen model of the IISM \citep{Williamson1972} to describe the scatter-broadening of the intrinsic pulse from the pulsar. The scattering timescale ($\tt \tau_{sc}$) is then calculated using
\begin{equation}
        \log(\tau_{sc}) = \log(\tau_{ref}) + \alpha \times \log(f) - \alpha \times \log (300) \, ,
\end{equation} 
where $f$ is the frequency, and  $\tt \tau_{ref}$ is the pulse broadening at the reference frequency of 300 MHz. 
As explained later, we used both a constant $\alpha$ (-4.4) assuming the Kolmogorov 
spectrum as well as a variable $\alpha$. 
The simulated pulse was then convolved with the pulse broadening function, $\exp(-\phi/\tau_{sc})$ for each 
sub-band. Next, we generated  the required noise for a given S/N as explained earlier and 
added this to the scattered pulse.

Then, we injected epoch to epoch DM variations using a DM time-series  as given below:
\begin{equation}
DM(t) = DM_0 + \delta DM(t - t_0)^3 , 
\end{equation}
where $DM_0$ is the fiducial DM at  $t_0$, chosen as the first 
epoch, over an observation interval spanning 10 years, sampled once every month. 
Three data sets with different amplitudes of DM variations, namely 0.01 (DMe-2), 0.001 (DMe-3), and 
0.0001 (DMe-4) pc\,cm$^{-3}$, were generated. A phase delay corresponding to the simulated 
DM at a given epoch was calculated with phase predictors using \texttt{TEMPO2} \citep{tempo2I} for each sub-band, and the simulated and scattered pulse was 
placed at this phase delay by shifting it by the calculated delay. Finally, this frequency resolved simulated 
data were written to an output PSRFITS file.

For each amplitude of the DM variation, three sets of simulated data were produced. The first set of simulated data had only the DM variation with no scatter-broadening (NS case). In the second set of simulated data, along with the DM variations, we also incorporated scatter-broadening effect with a constant value of the frequency scaling index, $\alpha = -4.4$, assuming a Kolmogorov turbulence (CS case). The value of $\tau$ at 300 MHz was chosen to be 0.7 ms. In the third set, along with the DM variations, we incorporated a variation in the frequency scaling index, $\alpha$ (VS  case). Here, we used the measurements of frequency scaling index, $\alpha$ for PSR J1643$-$1224 as the injected $\alpha$. The value of $\tau$ at 300 MHz was fixed for all the profiles and scaled accordingly with the frequency. 
Thus, we simulated 21 data-sets, each with 120 epochs, for the three different cases.
 
First, the simulated data-sets 
were  used to understand the effect of 
scatter-broadening on the estimates of DM. Then, our improved technique 
was tested and evaluated on the simulated data for 
CS and VS cases. The results of these analyses are presented 
in the following sections.

\subsection{Effect of scatter-broadening on DM Estimates}

We used \texttt{DMCalc} \citep{krishnakumar_et_al2021} on these simulated pulsar 
profiles to estimate the DMs for all the cases. In order to run \texttt{DMCalc}, 
we selected a high S/N ratio (from the 2000 S/N case) template for each case. The DMs were estimated for the  simulated data-sets spanning the range of S/N for all the three cases: NS, CS and VS. The results are presented in Figure~\ref{d4before}, where the plots of estimated DMs are shown 
after subtracting the injected DMs for simulated data-sets with S/N equal to 20 and 400,  and the amplitude of DM variations equal to  0.0001. The mean difference between the estimated and the injected DMs over all epochs 
and its standard deviation are also listed in the third and fifth columns of Table \ref{mean-sd}, respectively. 

As the pulse is without scatter-broadening in the NS case, the estimated DMs were consistent with the injected DMs for the full range of S/N, with the mean difference smaller than the DM uncertainty. In the CS and VS cases, where the simulated 
data-set consists of scatter-broadened pulse, the DMs were estimated with a bias, seen as offsets in Figure \ref{d4before} and significant mean difference in Table \ref{mean-sd}.  The bias is smaller for the VS case than for the CS case. 

The standard deviation in Table \ref{mean-sd} gives an idea of the 
variability in the DM estimates over all the epochs. The estimated DMs had larger variability for cases 
with S/N less than 50.  Another interesting feature in our results is that the variability was 
larger for the VS case as compared to the CS case, suggesting a larger fluctuation of DM estimates 
for pulsars showing variable scatter-broadening with observation epochs. These trends were consistent for all cases of DM variations.

\begin{figure}
 \centering
\includegraphics[scale=0.25,trim={0cm 0cm 0cm 0cm},clip]{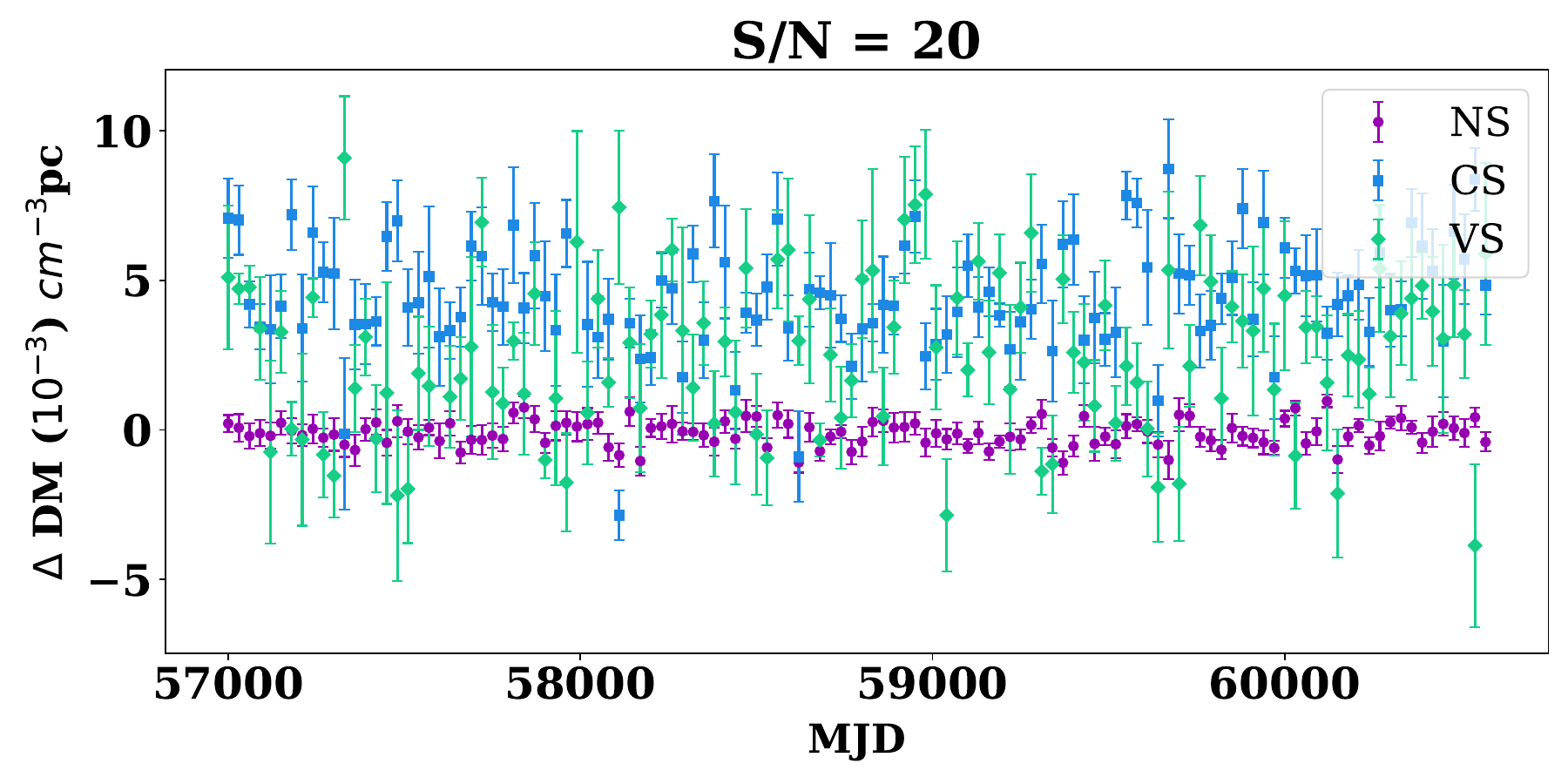}
\includegraphics[scale=0.25,trim={0cm 0cm 0cm 0cm},clip]{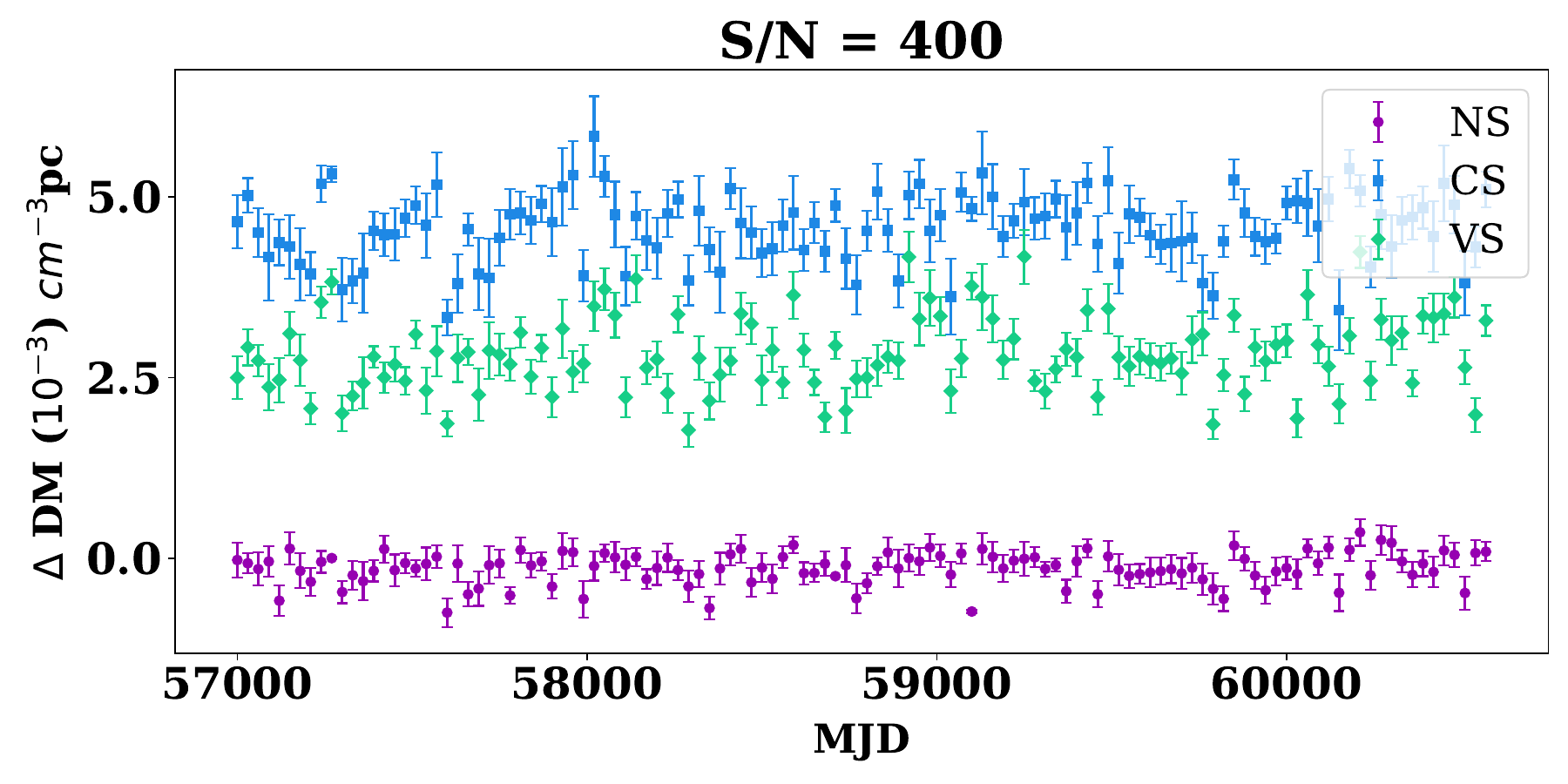}
\caption{The difference between injected and estimated DMs ($\Delta$ DM) for three cases: no scattering (NS), constant scattering (CS), and variable scattering (VS) for the set of files generated with S/N = 20 (upper panel) and 400 (lower panel) with injected DM variations of the order 0.0001 cm$^{-3}$~pc. These measurements were carried out on the simulated data before the application of \texttt{DMscat}.}
\label{d4before}
\end{figure}

\begin{figure}
 \centering
\includegraphics[scale=0.25,trim={0cm 0cm 0cm 0cm},clip]{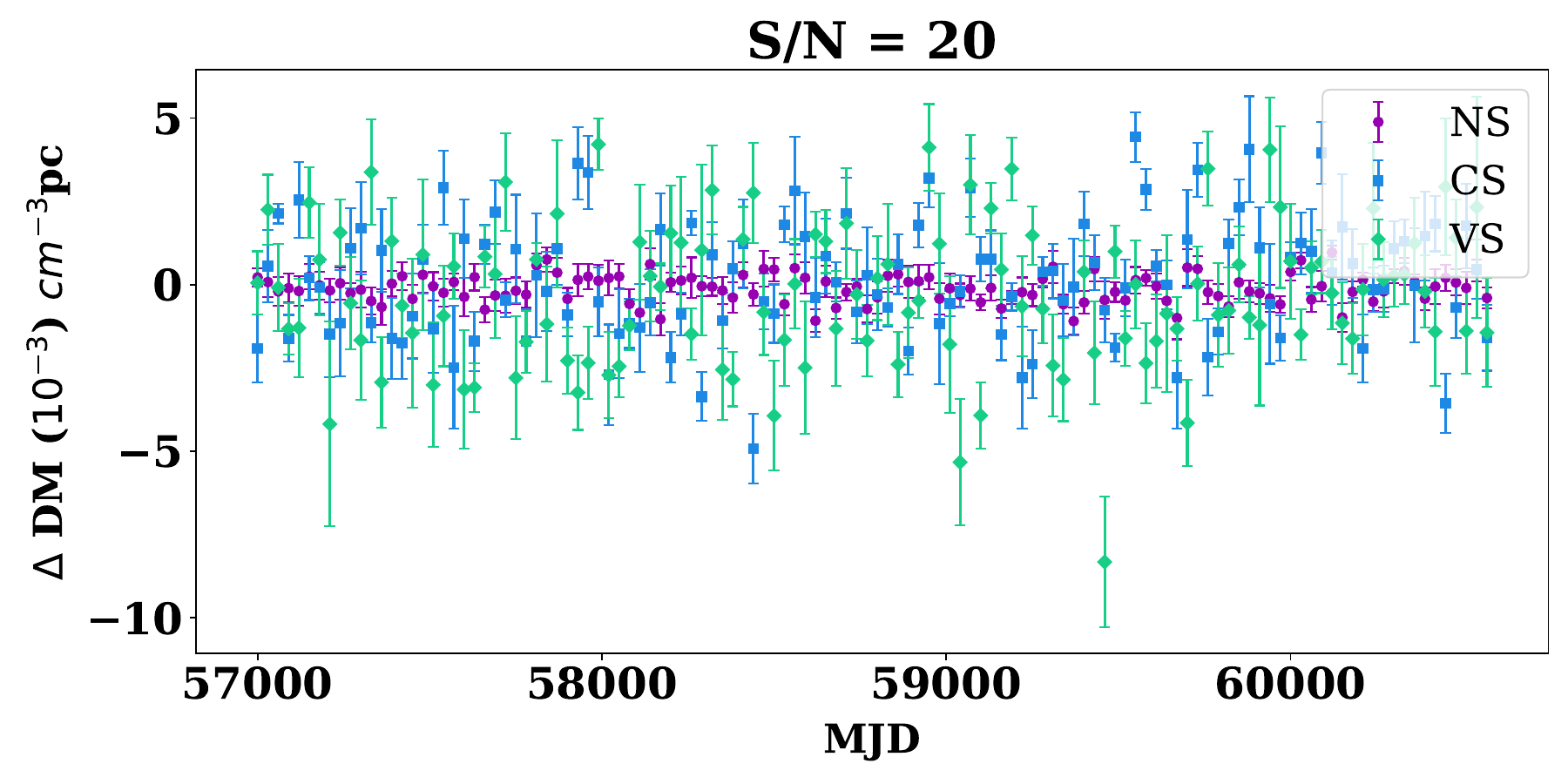}
\includegraphics[scale=0.25,trim={0cm 0cm 0cm 0cm},clip]{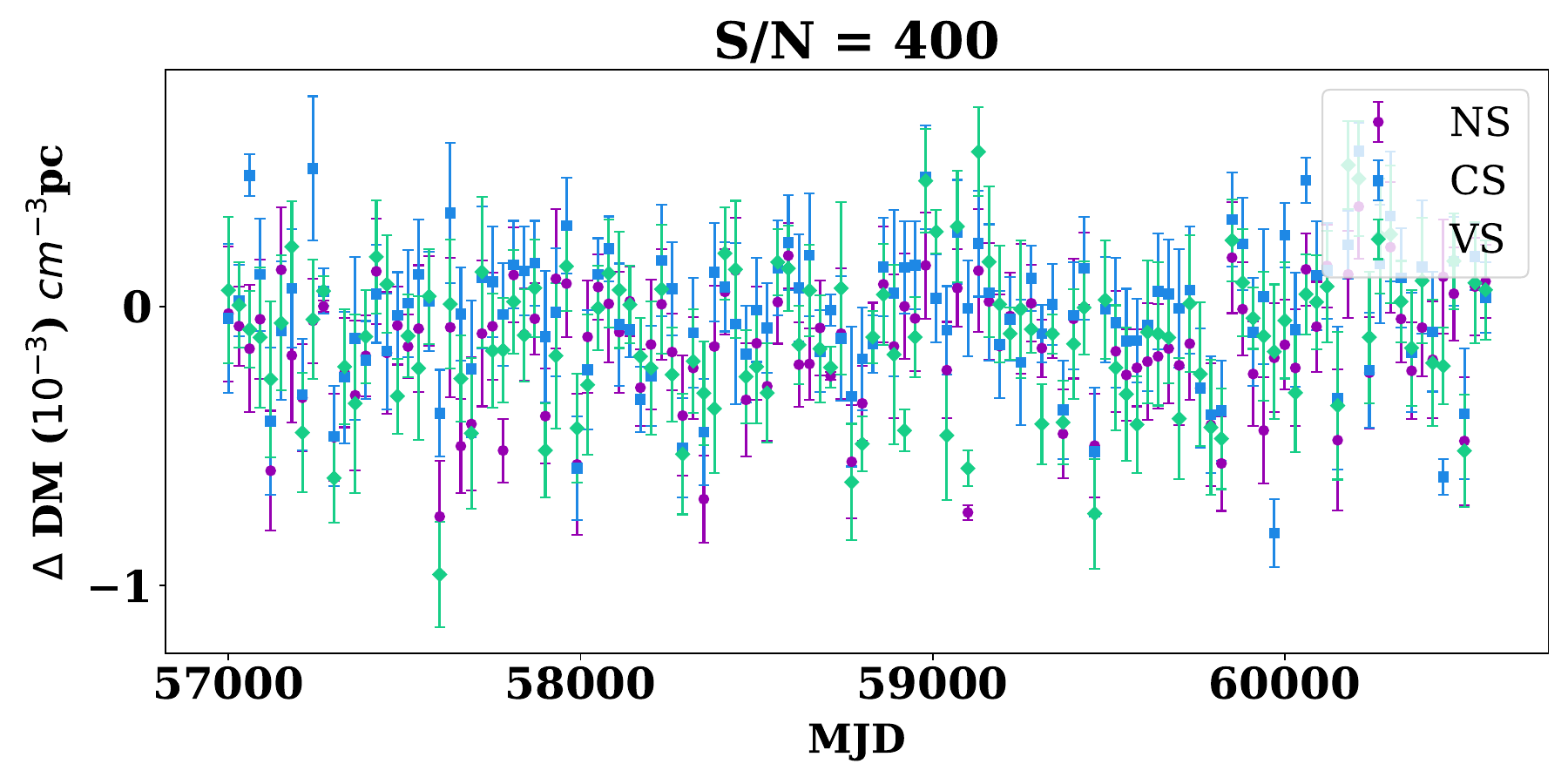}
\caption{The difference between the injected and estimated DMs ($\Delta$ DM) for three cases: no scattering (NS), constant scattering (CS), and variable scattering (VS) for the set of files generated with S/N = 20 (upper panel) and 400 (lower panel) with injected DM variations of the order 0.0001 cm$^{-3}$~pc. These measurements were carried out on the simulated data after the application of \texttt{DMscat} .} 
\label{d4after}
\end{figure}

\begin{figure}
 \centering
\includegraphics[scale=0.25,trim={0cm 0cm 0cm 0cm},clip]{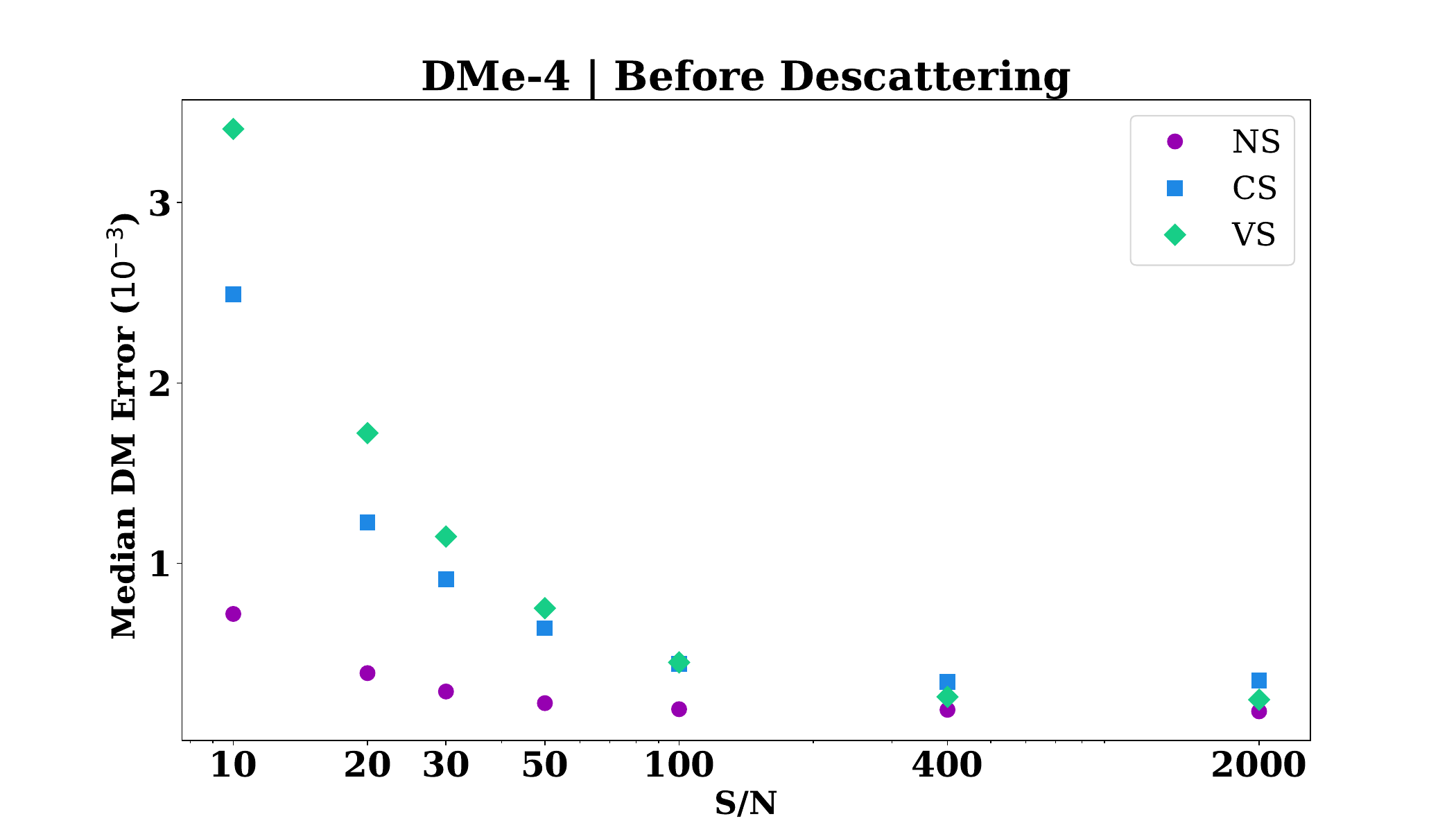}
\includegraphics[scale=0.25,trim={0cm 0cm 0cm 0cm},clip]{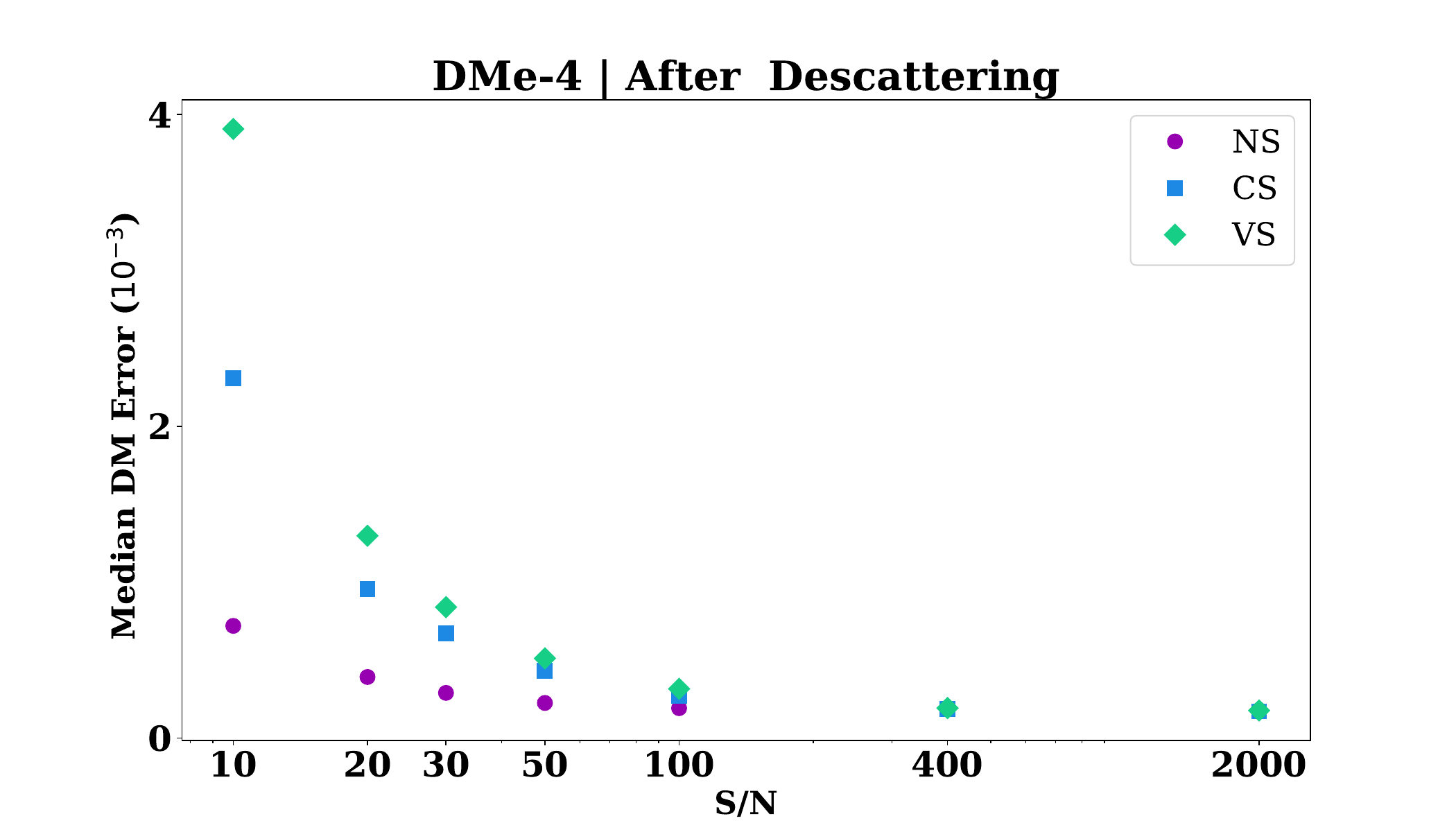}
\caption{The variation of median error in the DM estimation with respect to S/N before (upper panel) and after (lower panel) the application of \texttt{DMscat}. There is no precision in the injected DMs but the DMs vary in the order of $10^{-4}$ pc/cc. The median errors plotted here are an order of magnitude larger than the variation in the injected DMs. Please note that the errors here refer to precision in DM measurements and not the offset or bias in DM estimates.}
\label{dmesnr}
\end{figure}

\subsection{Testing \texttt{DMscat} on simulated data}
\label{sec:techtestsim}

We used the simulated data sets in order to demonstrate and test our method of removing the effect 
of  scatter-broadening on the DM estimates. We tested \texttt{DMscat} on the CS and VS datasets 
to generate new pulse profiles. First, we compared the recovered profiles for 
each sub-band against the injected profiles by subtracting the recovered profile 
from the injected profile. The obtained residuals were normally distributed
and consistent with the noise injected in the simulated data demonstrating 
that the technique works well on the simulated data, particularly for S/N greater than 100. The technique worked for both CS and VS cases for different DM variations. 

Then, we used \texttt{DMCalc} on these new profiles to estimate the DMs. The results are 
shown in Figure \ref{d4after}. We have plotted the difference between the estimated and  injected DMs, which is a measure of the bias in the DM measurements. These plots are shown for the synthetic data,  with the S/N equal to 20 and 400 and amplitude of DM variations equal to  0.0001 pc cm$^{-3}$. The mean difference between the estimated and the injected DMs over all epochs 
and their standard deviations are also collated in the fourth and sixth columns of Table \ref{mean-sd}. Broadly, the mean of the estimated DMs for the CS and VS cases 
were consistent with the ones obtained for the NS case, while the variability, reflected by the standard 
deviation was larger for CS and VS cases as compared to the NS case. 

The estimated and the injected DMs were consistent within the DM uncertainties for both CS and VS cases for all S/N cases, as is evident from Table \ref{mean-sd}, indicating that 
the technique is able to recover the injected DMs without the bias seen in the scatter-broadened 
data. Moreover, the variability of DM estimates over epochs is reduced by about half 
for S/N above 100, whereas the variability is the same or worse for S/N below 100. This validates \texttt{DMscat} 
and suggests that the technique will be useful in reducing the scatter-broadening noise for S/N 
larger than 100. It is important to emphasize that the technique reduces the bias in the estimates of DM,  but may not improve the corresponding precision, i.e., the uncertainties on DMs may not change in this procedure.

The two panels of Figure \ref{dmesnr} compare the median DM uncertainities before and after the application of \texttt{DMscat} for the case where the DM variation is of the order of $10^{-4}$ pc/cc. The median uncertainties are an order of magnitude larger than the variation in the injected DMs. After the application of \texttt{DMscat}, the median DM error is similar for NS, CS, and VS cases with S/N larger than 100,  whereas for data-sets with S/N lower than 100, the median error does not seem to improve. The aim of our technique is to obtain DM estimates closer to reality and this plot suggests that our technique significantly degrades the precision of DM at lower S/N. 

\begin{figure}
 \centering
\includegraphics[scale=0.52,trim={0cm 0cm 0cm 0cm},clip]{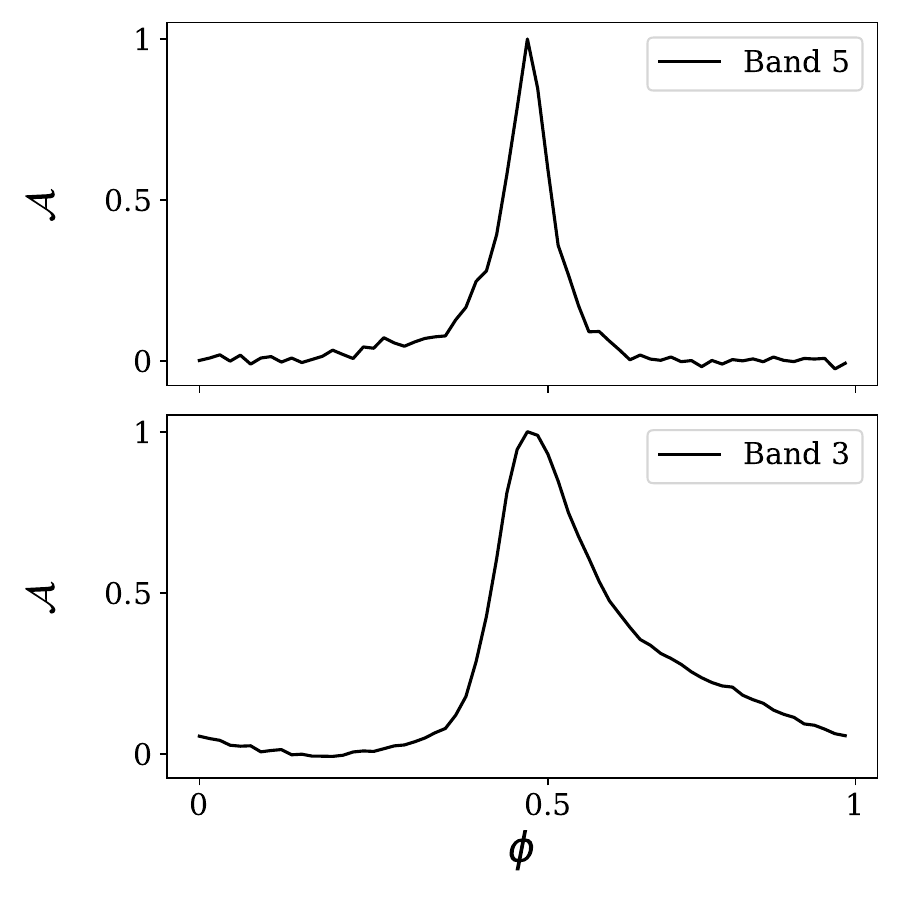}
\caption{Plots showing the scatter-broadening in PSR J1643$-$1224 in Band 3 (lower panel) and a sharp profile with negligible scattering in Band 5 (upper panel). Here $\mathcal{A}$ is the amplitude (in arbitrary units) and $\phi$ is the pulse phase.} 
\label{fig:1643figs}
\end{figure}

\begin{table*}
\centering
\caption{The values of mean and standard deviations of the differences of injected versus estimated DMs for various cases with different S/N for the simulations with DM variation of the order 0.0001.}
\label{mean-sd}
\setlength{\tabcolsep}{4pt}
 {\renewcommand\arraystretch{1.1}
\begin{tabular}{c|ccccc}
\hline \hline
\multirow{2}{*}{\textbf{S/N value}} & \multirow{2}{*}{\textbf{Cases}} &\multicolumn{2}{c}{\textbf{Mean (10$^{-3}$}) cm$^{-3}$ pc} & \multicolumn{2}{c}{\textbf{Standard Deviation (10$^{-3}$}) cm$^{-3}$ pc}\\  
   & & Before & After & Before & After\\  \hline
\multirow{3}{*}{\textbf{10}}  &  NS  &  0.015   &    &  0.37   &    \\
     &  CS  &   4.8  &   -0.46 &  4.5   &  4.6 \\
     &  VS  &  5.7   &  -2.4   &  10.5   & 11.6  \\ \hline
\multirow{3}{*}{\textbf{20}}   &  NS  &  -0.10   &     &  0.41 &  \\
     &  CS  &   4.4  & 0.16    &  1.8   & 1.7 \\
     &  VS  &  2.5   &  -0.31   &   2.5  &  2.1 \\ \hline 
\multirow{3}{*}{\textbf{30}}  &  NS  &  -0.11   &      &  0.30 & \\
     &  CS  &  4.3   &  -0.016   &   1.1  & 1.0 \\
     &  VS  &  2.4   &  -0.42   &  1.7   & 1.2 \\  \hline 
\multirow{3}{*}{\textbf{50}}  &  NS  &  -0.11   &     &  0.23   &  \\
     &  CS  &  4.3   &  -0.0032   &  0.69   &  0.63\\
     &  VS  &  2.5   &  -0.24   &  1.0   & 0.79 \\  \hline
\multirow{3}{*}{\textbf{100}}  &  NS  &  -0.13   &     &  0.22   &  \\
     &  CS  &   4.5  &   -0.0097  &   0.48  &  0.35\\
     &  VS  &   2.6  &   -0.17  &  0.68   &  0.43 \\ \hline 
\multirow{3}{*}{\textbf{400}}  &  NS  &  -0.13   &     &  0.21   &  \\
     &  CS  &  4.5   & -0.026    &  0.46   & 0.23 \\
     &  VS  &  2.8   & -0.11    &  0.51   & 0.25 \\  \hline 
\multirow{3}{*}{\textbf{2000}}  &  NS  &  -0.11   &     &   0.21  &  \\
     &  CS  &  4.5   &  -0.018   &  0.48   & 0.22 \\
     &  VS  &  2.8   &  -0.083   &  0.5   & 0.22 \\  \hline
\end{tabular}
}
\end{table*}

\begin{figure}
 \centering
\includegraphics[scale=0.45,trim={0cm 0cm 0cm 0cm},clip]{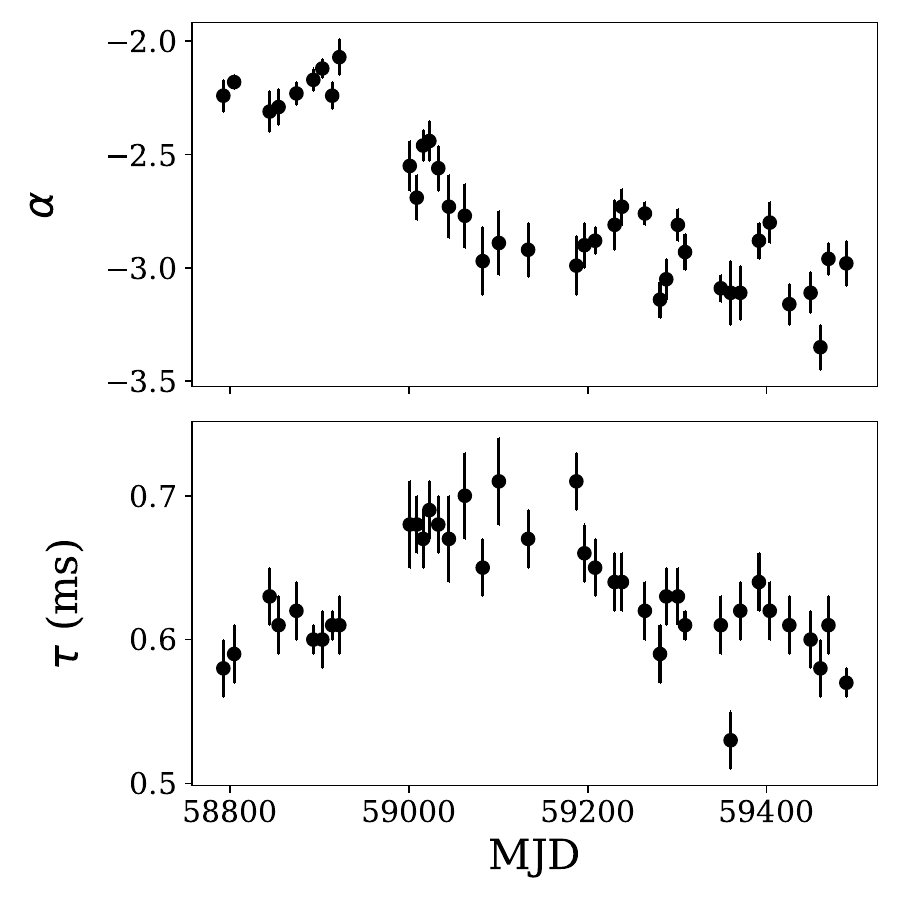}
\caption{Upper panel: The frequency scaling index ($\alpha$) over Band 3 is plotted in this figure for 
PSR J1643$-$1224 from 2019 to 2021. Lower panel: The estimated scatter-broadening time ($\tau_{sc}$ at 406 MHZ, near the band centre) 
data for PSR J1643$-$1224 is shown as a function of observing epoch. }
\label{fig:alphatau}
\end{figure}

\begin{figure*}
 \centering
\includegraphics[scale=0.35,trim={0cm 0cm 0cm 0cm},clip]{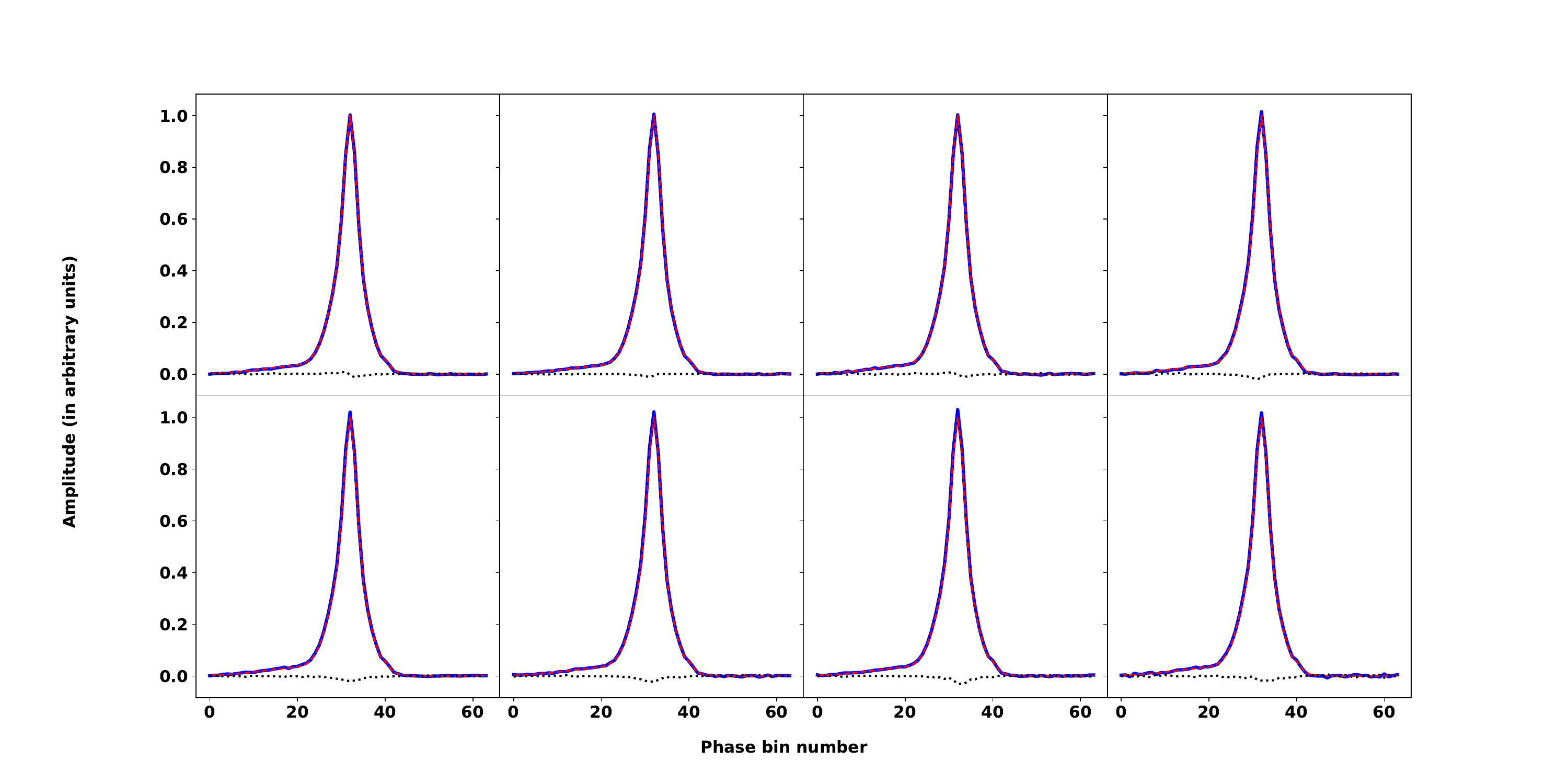}
\caption{The figure shows the comparison of the reconstructed profile in Band 3 with respect to the Band 5 template which was used to descatter the profile. The solid curve in blue indicates the Band 5 template, the red dashed curve shows the Band 3 profile after removing scattering, with the assumption that the profile does not evolve with frequency. The black dotted curve indicates the residuals, i.e., the difference between the template and the Band 3 profile at every frequency channel. }
\label{profreal}
\end{figure*}

\section{Application of \texttt{DMscat} on PSR J1643-1224} \label{sec:result_1643}

After validating \texttt{DMscat}, we applied this technique to PSR J1643$-$1224 data observed with the uGMRT as  part of the InPTA observations. PSR J1643$-$1224 is a pulsar in the PTA ensemble that exhibits prominent scatter-broadening. This 
pulsar is observed in the InPTA experiment simultaneously at two different frequency bands, namely Band 3 (300$-$500 MHz) and Band 5 (1260$-$1460 MHz), using the upgraded Giant Metrewave Radio Telescope~\citep{ugmrt, reddy_et_al2017}. The uGMRT observations at Band 3 and Band 5 are concurrent with all backend delays well calibrated and therefore such uGMRT data does not require a JUMP and can
directly be used to estimate DMs by combining the two bands \citep{inptadr1}. These simultaneous observations at two different bands allow us to estimate the DMs with high precision. 
Negligible scatter-broadening is seen in Band 5 data, whereas the pulsar shows significant pulse broadening 
at Band 3 as can be seen in Fig.~\ref{fig:1643figs}. We used  the observations over two years between 2019 and 2021, which also formed part of InPTA Data 
Release 1 \citep[InPTA-DR1 : ][]{inptadr1}. We only used the data observed with 200 MHz bandwidth (MJD 58781 $-$ 
59496). The DM time series of this pulsar, obtained with \texttt{DMCalc} using data without accounting for scatter-broadening, were presented in the InPTA-DR1 and is 
shown in Figure \ref{realdm}. 

First, the Band 5 data for PSR J1643$-$1224 were collapsed across the band to obtain a template for 
the highest S/N epoch (MJD 59308). This profile was further denoised using the \texttt{paas} program of \texttt{PSRCHIVE}. Such a template generated from single epoch high S/N profile observations  have previously  been used in InPTA-DR1 too \citep{inptadr1}. In our present case, we have further de-noised the high S/N profile to generate noise free template.  
This template can still be effected by the problem of self-noise as mentioned in \cite{Wang_2022, Hotan_2004_1022}. The final noise free template was used for all further analysis. Band 3 data were collapsed to 8 sub-bands. Then, we obtained the estimates of $\tau_{sc}$ for each of the 8 sub-bands 
and $\alpha$ as described in Section \ref{sec:technique}. These are presented in Figure~\ref{fig:alphatau}. Significant variations are seen in both the parameters over the two year time-scale of the data, which 
suggests that the  DM estimates are likely to have a time-varying bias due to scatter-broadening. This, 
coupled with epoch-dependent time delays due to scatter-broadening itself needs to be accounted for 
in this pulsar for a meaningful GW analysis. Further, the median frequency scaling index was estimated to be -2.84, which was different 
from Kolmogorov turbulence (-4.4). 

We used the estimates of $\tau_{sc}$ and $\alpha$ presented in Figures \ref{fig:alphatau} 
to remove scatter-broadening in the pulse using \texttt{DMscat} as explained in Section \ref{sec:technique}.
We show a comparison of the Band 3 reconstructed profiles at different frequency channels with the Band 5 template for MJD 58914 in Figure \ref{profreal}. The residuals obtained by subtracting the two profiles at every sub-band are also shown in this figure. Application of the  Anderson-Darling  test \citep{AndersonDarling_1952} shows
that these residuals were normally distributed. Therefore, \texttt{DMscat} is able to recover 
the profile without scatter-broadening. 

The resultant PSRFITS files were analysed with  \texttt{DMCalc}
to estimate the DMs. The estimated DMs after the application of \texttt{DMscat} are shown in Figure~\ref{realdm} 
along with the DM obtained in InPTA-DR1. Our technique takes care of scatter broadening and we believe that the DM series estimated with the reconstructed profiles are closer to reality and hence would differ slightly from the DM series presented in InPTA-DR1.

\begin{figure}
 \centering
\includegraphics[scale=0.26,trim={0cm 0cm 0cm 0cm},clip]{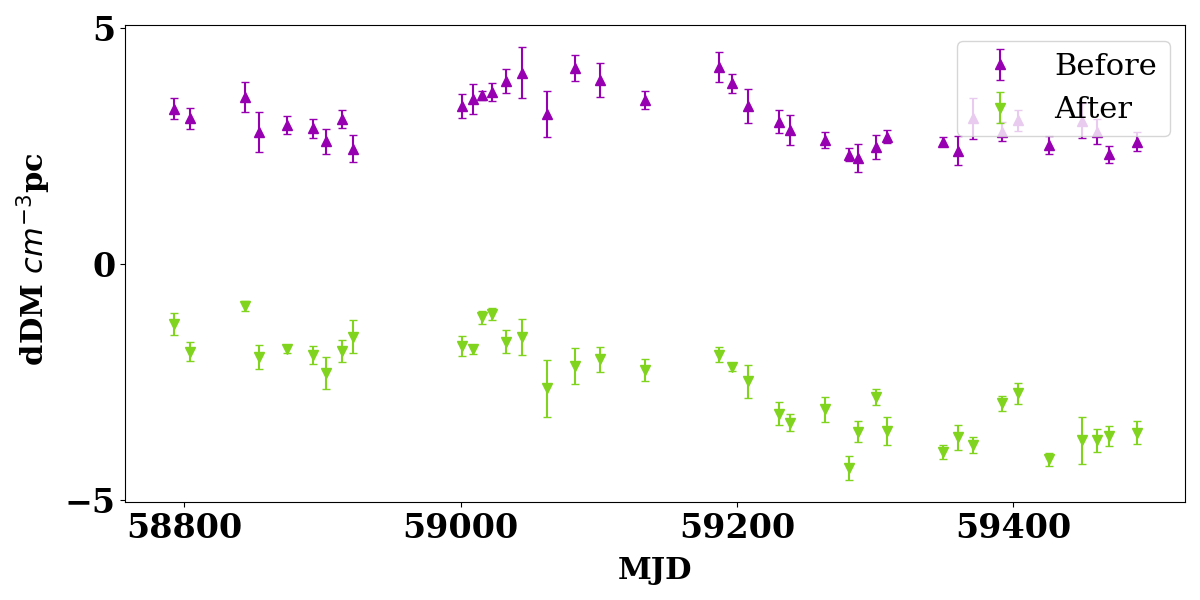}
\caption{The DM time series for the InPTA dataset of PSR J1643$-$1224 before and after the application of the technique. Here, dDM is the offset between the estimated and the fiducial DM used to align the template.} 
\label{realdm}
\end{figure}

\section{Conclusions} \label{sec:conclusion}
In this paper, we have demonstrated that the pulse-broadening in pulsar data can affect the estimates of 
DM using wide-band observations. Using simulated data, we show that a bias is seen in the DM estimates in 
scatter-broadened data. 
This bias depends on the spectral index of turbulence. 
The variability of the DM estimates over different epochs was found to be larger for scattering with a variable $\alpha$, suggesting that 
the DM noise estimates may be less reliable for scattering with a variable $\alpha$. An improved technique, \texttt{DMscat}, for removing the pulse-broadening due to multi-path propagation 
in the IISM is presented in this paper to remove the observed bias. The technique was validated with tests on simulated data, 
where it was shown that the estimated DMs are consistent with the injected ones. 
The technique will be useful in reducing 
the scattering noise for S/N larger than 100. The measurements of the frequency scaling index, $\alpha$, 
and scatter-broadening time, $\tau_{sc}$,  were presented for PSR J1643$-$1224 
observed using the uGMRT as part of the InPTA project. Both $\alpha$ and $\tau_{sc}$ were found to vary with observational epochs and $\alpha$ was measured to be different from that 
expected for a medium with Kolmogorov turbulence, which was also seen in \cite{Main_2023}. This could be due to the presence of the  HII region Sh 2-27 \citep{Mall_2022} in the line of sight. We will investigate this in a future work. \texttt{DMscat} was applied to 
PSR J1643$-$1224 to obtain a DM time-series from profiles without pulse-broadening.   
Thus, we have demonstrated the applicability of \texttt{DMscat} both on simulated data-sets and observed pulsar data under the assumption that there 
is negligible frequency evolution of the profile. 

A few pulsars among the PTA sample, such as PSRs J1643$-$1224 and J1939+2134, 
show significant DM variations as well as scatter-broadening at low frequencies. 
While these are bright pulsars with a high potential for precision timing, the 
variation in the ToA delays due to scattering most likely limits their contribution to a 
PTA experiment. Typically, such IISM variations are removed from timing residuals 
by modeling these chromatic noise sources as Gaussian processes (GP). In most of the 
recent PTA work, the IISM noise is modeled as a DM GP process with a $\nu^{-2}$ dependence~\citep{vanHaasteren2014, Lentati2014GP}.
The presence of scattering can lead to a leakage of the IISM noise in achromatic noise 
models, which can introduce subtle systematics in decade long PTA data-sets, particularly 
when the time scale of such chromatic variations are similar to achromatic or deterministic  
variations. An analysis after the application of \texttt{DMscat} can potentially help 
in the robust determination of these models at least for PSR J1643$-$1224. We intend to 
carry out such analysis as a follow-up work.

The main limitation of the method is that it may not work when frequency evolution 
of the  profiles is present. Techniques to address this limitation are motivated by 
this work. Possibilities are a modification of the  wide-band techniques \citep{Penucci_2014, Nobleson_2022, Paladi_2023}. Such developments are intended in the near future, which 
could be tested on the simulated data as well as actual observations. 

With the recently announced evidence favouring a spatially correlated signal linked to the gravitational wave background
\citep{Agazie_2023, Antoniadis_2023, Reardon_2023, Xu_2023}, the development of methods such as \texttt{DMscat} 
and others would possibly help in improving the significance in the upcoming IPTA Data Release 3. We shall investigate this possibility in a future work.

\section*{Acknowledgements}
InPTA acknowledges the support of the GMRT staff in resolving technical difficulties and providing technical solutions for high-precision work. We acknowledge the GMRT telescope operators for the observations. The GMRT is run by the National Centre for Radio Astrophysics of the Tata Institute of Fundamental Research, India.
JS acknowledges funding from the South African Research Chairs Initiative of the Depart of Science and Technology and the National Research Foundation of South Africa. 
BCJ acknowledges support from Raja Ramanna Chair (Track - I) grant from the Department of Atomic Energy, Government of India. BCJ also acknowledges support from the Department of Atomic Energy Government of India, under project number 12-R\&D-TFR-5.02-0700. DD and MB acknowledge the support from the Department of Atomic
Energy, Government of India through Apex Project - Advance Research
and Education in Mathematical Sciences at IMSc.
SD  and AS acknowledge the grant T-641 (DST-ICPS).
YG acknowledges support from the Department of Atomic Energy, Government of India, under project number 12-R\&D-TFR-5.02-0700.
TK is supported by the Terada-Torahiko Fellowship and the JSPS Overseas Challenge Program for Young Researchers.
AKP is supported by CSIR fellowship Grant number 09/0079(15784)/2022-EMR-I.
AmS is supported by CSIR fellowship Grant number 09/1001(12656)/2021-EMR-I and DST-ICPS T-641.
AS is supported by the NANOGrav NSF Physics Frontiers Center (Awards No 1430284 and 2020265).
KT is partially supported by JSPS KAKENHI Grant Numbers 20H00180, 21H01130, 21H04467 and JPJSBP        120237710, and the ISM Cooperative Research Program (2023-ISMCRP-2046).

\section*{Data Availability}
The data underlying this article will be shared on reasonable request to the corresponding author. 

\section*{Software}
\href{http://dspsr.sourceforge.net/}{\texttt{DSPSR}} \citep{DSPSR}, 
\href{http://psrchive.sourceforge.net/}{\texttt{PSRCHIVE}} \citep{Hotan+2004}, 
\href{https://github.com/ymaan4/RFIClean}{\texttt{RFIClean}} \citep{Maan+2020},
\href{https://github.com/abhisrkckl/pinta}{\texttt{PINTA}} \citep{pinta},
\href{https://bitbucket.org/psrsoft/tempo2/src/master/}{\texttt{TEMPO2}} \citep{tempo2I,Tempo2II}, 
\href{https://github.com/kkma89/dmcalc}{\texttt{DMCALC}} \citep{krishnakumar_et_al2021},
\href{https://github.com/lmfit}{\texttt{lmfit}} \citep{nsai2014},
\href{https://github.com/matplotlib/matplotlib}{\texttt{matplotlib}} \citep{h2007},
\href{https://astropy.org}{\texttt{astropy}} \citep{astropy+2018}.



\bibliographystyle{mnras}
\bibliography{reference} 

\section*{Affiliations}
\noindent
{\it 
$^{1}$High Energy Physics, Cosmology \& Astrophysics Theory (HEPCAT) Group,
Department of Mathematics and Applied Mathematics,\\
University of Cape Town, Cape Town 7700, South Africa \\
$^{2}$Department of Physics, Indian Institute of Technology Roorkee, Roorkee 247667, Uttarakhand, India\\
$^{3}$National Centre for Radio Astrophysics, Tata Institute of Fundamental Research, Pune 411007, Maharashtra, India\\
$^4$Max-Planck-Institut f{\"u}r Radioastronomie, Auf dem H{\"u}gel 69, 53121 Bonn, Germany\\
$^5$Fakult{\"a}t f{\"u}r Physik, Universit{\"a}t Bielefeld, Postfach 100131, 33501 Bielefeld, Germany\\
$^6$Department of Physical Sciences, IISER Kolkata, West Bengal, India \\
$^7$Center of Excellence in Space Sciences India, IISER Kolkata, West Bengal, India \\
$^8$Indian Institute of Science Education and Research, Mohali, 140306, India\\
$^9$Department of Earth and Space Sciences, Indian Institute of Space Science and Technology, Valiamala P.O., Thiruvananthapuram 695547, Kerala, India\\
$^{10}$Department of Physics, Government Brennen College, Thalassery, Kannur University, Kannur 670106, Kerala, India\\
$^{11}$Department of Physics, IIT Hyderabad, Kandi, Telangana 502284\\
$^{12}$The Institute of Mathematical Sciences, C. I. T. Campus, Taramani, Chennai 600113, India\\ 
$^{13}$Department of Electrical Engineering, IIT Hyderabad, Kandi, Telangana 502284, India\\
$^{14}$Homi Bhabha National Institute, Training School Complex, Anushakti Nagar, Mumbai 400094, India\\
$^{15}$Department of Physics and Astrphysics, Delhi University, Delhi, India\\
$^{16}$Department of Astronomy and Astrophysics, Tata Institute of Fundamental Research, Homi Bhabha Road, Navy Nagar, Colaba, Mumbai 400005, India\\
$^{17}$International Research Organization for Advanced Science and Technology,
Kumamoto University, 2-39-1 Kurokami, Kumamoto 860-8555, Japan\\
$^{18}$Osaka Central Advanced Mathematical Institute, Osaka Metropolitan University, Osaka, 5588585, Japan\\
$^{19}$Faculty of Advanced Science and Technology,
Kumamoto University, 2-39-1 Kurokami, Kumamoto 860-8555, Japan\\
$^{20}$Kumamoto University, Graduate School of Science and Technology, Kumamoto, 860-8555, Japan\\
$^{21}$Australia Telescope National Facility, CSIRO, Space and Astronomy, PO Box 76, Epping, NSW 1710, Australia\\
$^{22}$Joint Astronomy Programme, Indian Institute of Science, Bengaluru, Karnataka, 560012, India\\
$^{23}$Raman Research Institute, Bangalore, 560080, India\\
$^{24}$Department of Physics, Indian Institute of Science Education and Research Bhopal, Bhopal Bypass Road, Bhauri, Bhopal 462 066, Madhya Pradesh, India\\
$^{25}$Center for Gravitation Cosmology and Astrophysics, University of Wisconsin-Milwaukee, Milwaukee, WI 53211, USA
}




\appendix


\bsp	
\label{lastpage}
\end{document}